\documentclass[english]{revtex4-1}
\usepackage[T1]{fontenc}
\usepackage[latin9]{inputenc}
\usepackage{amsmath}
\usepackage{amssymb}
\usepackage{graphicx}

\makeatletter
\numberwithin{equation}{section}

\usepackage{babel}

\makeatother

\usepackage{babel}
\begin{document}

\title{Non-classical behaviour of coherent states for systems constructed
using exceptional orthogonal polynomials}

\author{Scott E. Hoffmann\textsuperscript{1}, Véronique Hussin\textsuperscript{2},
Ian Marquette\textsuperscript{1} and Yao-Zhong Zhang\textsuperscript{1}}
\email{scott.hoffmann@uqconnect.edu.au}

\address{\textsuperscript{1}School of Mathematics and Physics, The University
of Queensland, Brisbane, QLD, 4072, Australia~\\
~\\
\textsuperscript{2}Département de Mathématiques et de Statistique,
Université de Montréal, Montréal, Québec, H3C 3J7, Canada}
\begin{abstract}
We construct the coherent states and Schrödinger cat states associated
with new types of ladder operators for a particular case of a rationally
extended harmonic oscillator involving type III Hermite exceptional
orthogonal polynomials. In addition to the coherent states of the
annihilation operator, $c,$ we form the linearised version, $\tilde{c},$
and obtain its coherent states. We find that while the coherent states
defined as eigenvectors of the annihilation operator $c$ display
only quantum behaviour, those of the linearised version, $\tilde{c},$
have position probability densities displaying distinct wavepackets
oscillating and colliding in the potential. The collisions are certainly
quantum, as interference fringes are produced, but the remaining evolution
indicates a classical analogue.
\end{abstract}
\maketitle

\section{Introduction}

Coherent states are of widespread interest \cite{Glauber1963a,Glauber1963b,Klauder1963a,Klauder1963b,Barut1971,Perelomov1986,Gazeau1999,Ali2000,Quesne2001,Fernandez1995,Fernandez1999,GomezUllate2014}
because they can, in some cases, be the quantum-mechanical states
with the most classical behaviour. For the harmonic oscillator, the
energy eigenvectors have stationary position probability densities.
In contrast, the coherent state vectors, superpositions over all energy
eigenvectors, are represented by wavepackets that oscillate back and
forth in the $x^{2}$ potential, keeping the same Gaussian profile
at all times \cite{Cohen-Tannoudji1977}. For the free electromagnetic
field, the expectations of the field strength operators vanish in
state vectors with a definite number of photons. In the coherent state
superpositions over all photon number states, the field strengths
have nonzero expectations that behave classically and obey the free
Maxwell equations \cite{Glauber1963b}.

The coherent states of the isospectral (or almost isospectral) superpartners
of the harmonic oscillator have been studied by several authors, with
two classes of ladder operators having been used in the construction
\cite{Fernandez1994}. The first class is the \textit{natural}, of
the form
\begin{align}
b & =A\,a\,A^{\dagger},\quad b^{\dagger}=A\,a^{\dagger}A^{\dagger},\label{eq:0.1}
\end{align}
where $A$, $A^{\dagger}$ are the supercharges and $a,$ $a^{\dagger}$
are the ladder operators of the harmonic oscillator. The second class
is the \textit{intrinsic} class, that is a linearized version of the
natural, with ladder operators of the form
\begin{align}
\tilde{b} & =f(H^{(-)})b,\nonumber \\
\tilde{b}^{\dagger} & =b^{\dagger}f(H^{(-)}),\label{eq:0.2}
\end{align}
where $f$ is chosen so that these ladder operators then satisfy the
Heisenberg algebra, like the operators $a$ and $a^{\dagger}$. Here
$H^{(-)}$ is the supersymmetric partner Hamiltonian of the harmonic
oscillator Hamiltonian $H^{(+)}.$ Note that a relation of the form
of Eq. (\ref{eq:0.2}), with general function $f,$ was recently used
to define nonlinear supercoherent states \cite{DiazBautista2015}.

Other types of ladder operators have been discussed in connection
with Painlevé IV transcendent systems. Authors in \cite{Bermudez2015}
constructed a deformation of the harmonic oscillator that obeys a
second order polynomial Heisenberg algebra and the Painlevé IV equation.
They constructed the Barut-Girardello coherent states and obtained
results for the energy expectations. A class of deformations of the
harmonic oscillator called generalized isotonic oscillators have also
been studied by many authors. In particular coherent states were constructed
for natural and intrinsic ladder operators, using the Barut-Girardello
and the displacement operator definitions, and non-classical behaviour
was observed \cite{Ruby2010}.

All of these systems are in fact part of more general families of
deformations of the harmonic oscillator connected with the recently
discovered Hermite exceptional orthogonal polynomials \cite{Marquette2013b,Marquette2014b,Marquette2016}.
In this work we will focus on the existence of new ladder operators
\cite{Marquette2013b,Marquette2014b,Marquette2016}, outside the natural
and intrinsic classes, that connect the basis vectors and provide
the infinite-dimensional representations of a polynomial Heisenberg
algebra (no singlet ground state). These can be constructed via combinations
of supercharges of Darboux-Crum and Krein-Adler type. These are not
unique ladder operators connecting the basis states, in that they
exist alongside the natural and intrinsic operators, since there is
more than one path connecting the (+) state vectors (the eigenvectors
of $H^{(+)}$) and the (-) state vectors (the eigenvectors of $H^{(-)}$).

The system considered in this paper is a deformation of the harmonic
oscillator obtained with supersymmetric quantum mechanics (SUSY QM)
using Hermite exceptional orthogonal polynomials (Hermite EOPs) \cite{Marquette2013b}.
The ground state wavefunction of the partner Hamiltonian, $H^{(-)},$
is 
\begin{equation}
\phi(x)=(\frac{8}{\sqrt{\pi}})^{\frac{1}{2}}\frac{e^{-x^{2}/2}}{\mathcal{H}_{2}(x)}.\label{eq:1}
\end{equation}
Here the modified Hermite polynomials, real for all $m$ and positive
definite for even $m,$ are defined by 
\begin{align}
\mathcal{H}_{m}(x) & =(-i)^{m}H_{m}(ix).\label{eq:2.1}
\end{align}
The first three are
\begin{equation}
\mathcal{H}_{0}(x)=1,\quad\mathcal{H}_{1}(x)=2x,\quad\mathcal{H}_{2}(x)=4x^{2}+2.\label{eq:2.1.5}
\end{equation}
This choice (Eq.~(\ref{eq:1})) then determines the partner potential
and all other energy eigenvectors of the (-) system.

Note that this choice is the $m=2$ case of a class of Hamiltonians
using Hermite polynomials of order $m$ (even) \cite{Marquette2013b}.
Note also that $\mathcal{H}_{2}(x)$ is positive definite, so no singularity
is introduced, and $\phi(x)$ is square-integrable. We will see that
the partner potential for the (-) system is 
\begin{align}
V^{(-)}(x) & =x^{2}-2[\frac{\mathcal{H}_{2}^{\prime\prime}}{\mathcal{H}_{2}}-(\frac{\mathcal{H}_{2}^{\prime}}{\mathcal{H}_{2}})^{2}+1]=x^{2}+\frac{16(4x^{2}-2)}{(4x^{2}+2)^{2}}-2,\label{eq:2.5}
\end{align}
which displays (in Figure 1) a deep, narrow well around the origin.
This feature will dominate the physics of the system.

\begin{figure}
\begin{centering}
\includegraphics[width=6cm]{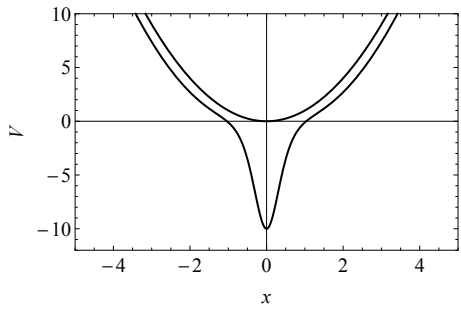}
\par\end{centering}
\caption{The partner potential $V^{(-)}(x)$ compared to $x^{2}.$}
\end{figure}

The focus of this paper will be the new ladder operators constructed
in \cite{Marquette2013b}, which we will denote by $c$ and $c^{\dagger}.$
The action of the raising operator, $c^{\dagger},$ is to increase
the index of the state vector by $3.$ So this can be used to form
three distinct, orthogonal ladders with the operators acting on three
orthogonal, disconnected, subspaces. One of these ladders includes
the ground state, given in Eq.~(\ref{eq:1}). Note that for the ladder
operators $b,b^{\dagger}$ and $\tilde{b},\tilde{b}^{\dagger}$ mentioned
above, the ground state is a singlet excluded from the ladders.

For comparison, we will also construct the operators $\tilde{c}$
and $\tilde{c}^{\dagger},$ which are linearized versions of $c$
and $c^{\dagger},$ respectively.

In this paper we will construct three distinct, orthogonal coherent
states of the $c$ operator, defined, for each complex value of $z,$
by 
\begin{equation}
c\,|\,z,c(\mu)\,\rangle=z\,|\,z,c(\mu)\,\rangle.\label{eq:3}
\end{equation}
Here $\mu=-3,1,2$ is the index of the lowest weight state vector
in the superposition. Note that this is not the only way to define
coherent states (see the discussion in Section IV below).

Similarly, we will construct the three sets of coherent states that
are eigenvectors, with complex eigenvalue $z,$ of the $\tilde{c}$
operator, defined by
\begin{equation}
\tilde{c}\,|\,z,\tilde{c}(\mu)\,\rangle=z\,|\,z,\tilde{c}(\mu)\,\rangle,\label{eq:4}
\end{equation}
for $\mu=-3,1,2.$

We will construct the coherent states associated with the new ladder
operator, $c,$ and with the linearized version, $\tilde{c}$, to
investigate their physical properties, to compare between the two
cases and to compare with the properties of the coherent states of
the $a$ operator.

This paper is organized as follow. In Section II we display the eigenfunctions
and spectra of the partner Hamiltonian for our system. In Section
III we construct two distinct types of ladder operator, the pair $c,c^{\dagger}$
and the linearized pair $\tilde{c},\tilde{c}^{\dagger}.$ In Section
IV we construct the coherent states corresponding to two different
annihilation operators for three cases each. In Section V we investigate
the properties of these coherent states. We plot the expectations
of total energy as functions of $|z|.$ Then we show the time-dependent
position probability densities for particular choices of the parameter
$z.$ Lastly, we show the time-dependent probability densities for
even and odd Schrödinger cat states formed from our coherent states.
In all of these cases, we compare with the familiar harmonic oscillator
coherent states. Conclusions follow in Section VI.

\section{Construction and Solution of the Partner Hamiltonians}

We use the scaling scheme $\hbar\rightarrow1,m_{0}\rightarrow1/2,\omega\rightarrow2,$
where $m_{0}$ is the mass of the particle and $\omega$ is the angular
frequency of the oscillator. Then the harmonic oscillator Hamiltonian
becomes 
\begin{equation}
H^{(0)}=-\frac{d^{2}}{dx^{2}}+x^{2},\label{eq:6}
\end{equation}
with energy spectrum 
\begin{equation}
E_{\nu}^{(0)}=2\nu+1,\quad\nu=0,1,2,\dots\label{eq:7}
\end{equation}
and normalized position eigenfunctions 
\begin{equation}
\psi_{\nu}^{(0)}(x)=\frac{1}{\sqrt{\sqrt{\pi}\,2^{\nu}\nu!}}e^{-x^{2}/2}H_{\nu}(x)\quad\mathrm{for}\ \nu=0,1,2,\dots\label{eq:8}
\end{equation}
satisfying
\begin{equation}
H^{(0)}\psi_{\nu}^{(0)}(x)=E_{\nu}^{(0)}\psi_{\nu}^{(0)}(x)\quad\mathrm{for}\ \nu=0,1,2,\dots.\label{eq:8.5}
\end{equation}

The creation and annihilation operators are, respectively, 
\begin{align}
a^{\dagger} & =-\frac{d}{dx}+x,\quad a=+\frac{d}{dx}+x,\label{eq:9}
\end{align}
satisfying the commutation relation 
\begin{equation}
[a,a^{\dagger}]=2.\label{eq:11}
\end{equation}
So the Hamiltonian can be written 
\begin{equation}
H^{(0)}=a^{\dagger}a+1.\label{eq:12}
\end{equation}

We summarize the SUSY QM construction of the partner Hamiltonians
$H^{(+)}$ and $H^{(-)}$ and their solutions \cite{Marquette2013a}.
The choice (\ref{eq:1}) of the ground state leads to the supercharges
\begin{align}
A & =+\frac{d}{dx}-\frac{\phi^{\prime}}{\phi}=+\frac{d}{dx}+x+\frac{\mathcal{H}_{2}^{\prime}}{\mathcal{H}_{2}},\nonumber \\
A^{\dagger} & =-\frac{d}{dx}-\frac{\phi^{\prime}}{\phi}=-\frac{d}{dx}+x+\frac{\mathcal{H}_{2}^{\prime}}{\mathcal{H}_{2}}.\label{eq:15}
\end{align}
Then the partner Hamiltonians are 
\begin{align}
H^{(+)} & =A^{\dagger}A=-\frac{d^{2}}{dx^{2}}+x^{2}+5,\nonumber \\
H^{(-)} & =AA^{\dagger}=-\frac{d^{2}}{dx^{2}}+x^{2}+\frac{16(4x^{2}-2)}{(4x^{2}+2)^{2}}+3.\label{eq:19}
\end{align}

The Hamiltonian $H^{(+)}$ is just the harmonic oscillator Hamiltonian
with a constant energy shift. So its spectrum is 
\begin{equation}
E_{\nu}^{(+)}=2(\nu+3)\quad\mathrm{for}\ \nu=0,1,2,\dots\label{eq:21}
\end{equation}
and its energy eigenfunctions are given by (\ref{eq:8}), $\psi_{\nu}^{(+)}(x)=\psi_{\nu}^{(0)}(x).$

By supersymmetry the Hamiltonian $H^{(-)}$ has a set of energy eigenvectors
with the same energies as those of $H^{(+)}$ and, in addition, a
ground state with zero energy. So we can write the spectrum as 
\begin{equation}
E_{\nu}^{(-)}=2(\nu+3)\quad\mathrm{for}\ \nu=-3,0,1,2,\dots\label{eq:22}
\end{equation}
The energy eigenfunctions of $H^{(-)}$ are 
\begin{equation}
\psi_{\nu}^{(-)}(x)=\begin{cases}
\sqrt{\frac{8}{\sqrt{\pi}}}\frac{e^{-x^{2}/2}}{\mathcal{H}_{2}(x)}y_{0}(x) & \mathrm{for}\ \nu=-3,\\
\frac{1}{\sqrt{\sqrt{\pi}2^{\nu+1}(\nu+3)\nu!}}\frac{e^{-x^{2}/2}}{\mathcal{H}_{2}(x)}y_{\nu+3}(x) & \mathrm{for}\ \nu=0,1,2,\dots
\end{cases},\label{eq:23}
\end{equation}
where 
\begin{align}
y_{0}(x) & =1,\nonumber \\
y_{\nu+3}(x) & =-\mathcal{H}_{2}(x)H_{\nu+1}(x)-4\mathcal{H}_{1}(x)H_{\nu}(x)\quad\mathrm{for}\ \nu=0,1,2,\dots\label{eq:25}
\end{align}
These functions, $y_{k}(x),$ are the Hermite exceptional orthogonal
polynomials (Hermite EOPs).

Many of the models for which coherent states have been constructed
and studied, such as the harmonic oscillator, the Morse potential
\cite{Angelova2012}, the Scarf potential \cite{Cooper1995} and the
infinite well \cite{Fiset2015}, admit the classical analogues of
ladder operators and are exactly solvable in classical mechanics \cite{Kuru2007}.

\section{Ladder Operators}

For the three categories of annihilation operators in the set $\{a,c,\tilde{c}\}$
and their Hermitian conjugates, we show the definitions and matrix
elements of the ladder operators and their (polynomial) Heisenberg
algebras. In the $c$ and $\tilde{c}$ cases we will find patterns
of zero modes that will break the spaces into distinct, orthogonal,
subspaces.

\subsection{The ladder operators $c$ and $c^{\dagger}$}

We briefly review the properties of the $a$ and $a^{\dagger}$ operators,
for comparison. The annihilation operator, $a,$ and the creation
operator, $a^{\dagger},$ defined by Eq.~(\ref{eq:9}), are ladder
operators connecting all of the eigenvectors $|\,\nu\,(+)\,\rangle$
of $H^{(+)}$ according to 
\begin{align}
a\,|\,\nu\,(+)\,\rangle & =\sqrt{2\nu}\,|\,\nu-1\,(+)\,\rangle,\quad a^{\dagger}\,|\,\nu\,(+)\,\rangle=\sqrt{2(\nu+1)}\,|\,\nu+1\,(+)\,\rangle,\quad\mathrm{for}\ \nu=0,1,2,\dots\label{eq:26}
\end{align}
With (from Equations (\ref{eq:19}),(\ref{eq:6}) and (\ref{eq:12}))
\begin{equation}
H^{(+)}=a^{\dagger}a+6,\label{eq:27}
\end{equation}
the ladder operators obey the Heisenberg algebra
\begin{align}
[H^{(+)},a] & =-2a,\quad[H^{(+)},a^{\dagger}]=+2a^{\dagger},\quad a^{\dagger}a=H^{(+)}-6,\quad a\,a^{\dagger}=H^{(+)}-4.\label{eq:28.1}
\end{align}
For these eigenvectors, the position wavefunctions are $\langle\,x\,|\,\nu\,(+)\,\rangle=\psi_{\nu}^{(+)}(x)=\psi_{\nu}^{(0)}(x),$
given by Eq.~(\ref{eq:8}). For all the following cases they are
$\langle\,x\,|\,\nu\,(-)\,\rangle=\psi_{\nu}^{(-)}(x),$ given by
Eq.~(\ref{eq:23}).

In \cite{Marquette2013b} new ladder operators are obtained, not of
the natural or intrinsic classes, for the $H^{(-)}$ eigenvectors.
These include nonzero matrix elements between the ground state and
the first excited state. This is possible because more than one path
(the other being the path for $b$ and $b^{\dagger}$) can be constructed
connecting the $H^{(+)}$ and $H^{(-)}$ systems. This, in turn, is
possible because of properties of the wavefunctions for the $H^{(-)}$
system, which are type III Hermite exceptional orthogonal polynomials.

They define first-order supercharges using Darboux-Crum and Krein-Adler
terms \cite{Marquette2013b},
\begin{align}
A_{1} & =+\frac{d}{dx}+x+\frac{\mathcal{H}_{0}^{\prime}}{\mathcal{H}_{0}}-\frac{\mathcal{H}_{1}^{\prime}}{\mathcal{H}_{1}},\quad A_{1}^{\dagger}=-\frac{d}{dx}+x+\frac{\mathcal{H}_{0}^{\prime}}{\mathcal{H}_{0}}-\frac{\mathcal{H}_{1}^{\prime}}{\mathcal{H}_{1}},\nonumber \\
A_{2} & =+\frac{d}{dx}+x+\frac{\mathcal{H}_{1}^{\prime}}{\mathcal{H}_{1}}-\frac{\mathcal{H}_{2}^{\prime}}{\mathcal{H}_{2}},\quad A_{2}^{\dagger}=-\frac{d}{dx}+x+\frac{\mathcal{H}_{1}^{\prime}}{\mathcal{H}_{1}}-\frac{\mathcal{H}_{2}^{\prime}}{\mathcal{H}_{2}}.\label{eq:53}
\end{align}
Then the product $A_{2}A_{1}$ maps from the $H^{(+)}$ eigenvectors
to the $H^{(-)}$ eigenvectors in $2$ steps and the operator 
\begin{equation}
c=A_{2}A_{1}A^{\dagger}\label{eq:56}
\end{equation}
acts as a lowering operator (in steps of $3$) in the $H^{(-)}$ system.
Likewise 
\begin{equation}
c^{\dagger}=AA_{1}^{\dagger}A_{2}^{\dagger}\label{eq:57}
\end{equation}
acts as a raising operator in the $H^{(-)}$ system. The operators
$c,c^{\dagger}$ obey the cubic polynomial Heisenberg algebra
\begin{align}
[H^{(-)},c] & =-6c,\quad[H^{(-)},c^{\dagger}]=+6c^{\dagger},\quad[c,c^{\dagger}]=Q(H^{(-)}+6)-Q(H^{(-)}),\label{eq:61.1}
\end{align}
where
\begin{equation}
Q(x)=x(x-8)(x-10).\label{eq:61.3}
\end{equation}
Their actions on the eigenvectors $|\,\nu\,(-)\,\rangle$ of $H^{(-)}$
are given by 
\begin{align}
c\,|\,\nu\,(-)\,\rangle & =-[2(\nu-1)2(\nu-2)2(\nu+3)]^{\frac{1}{2}}\,|\,\nu-3\,(-)\,\rangle,\nonumber \\
c^{\dagger}\,|\,\nu\,(-)\,\rangle & =-[2(\nu+2)2(\nu+1)2(\nu+6)]^{\frac{1}{2}}\,|\,\nu+3\,(-)\,\rangle.\label{eq:59}
\end{align}
As noted above, there are nonzero matrix elements of the form 
\begin{equation}
\langle\,0\,(-)\,|\,c^{\dagger}\,|\,-3\,(-)\,\rangle\neq0.\label{eq:60}
\end{equation}
Note that 
\begin{equation}
c\,|\,\mu\,(-)\,\rangle=0\quad\mathrm{for}\ \mu=-3,1,2.\label{eq:61}
\end{equation}
So there are three independent, mutually orthogonal, ladder subspaces
in this system, with lowest weights $\mu=-3,1,2.$ In the next section
we will thus construct three independent and mutually orthogonal coherent
states (for each complex value of $z$).

\subsection{The ladder operators $\tilde{c}$ and $\tilde{c}^{\dagger}$}

As discussed earlier, we linearize the operators $c$ and $c^{\dagger}$
to produce $\tilde{c}$ and $\tilde{c}^{\dagger}.$ First we relabel
the basis vectors:
\begin{equation}
|\,\kappa,\mu\,\rangle\equiv|\,\mu+3\kappa\,(-)\,\rangle\quad\mathrm{for}\ \kappa=0,1,2,\dots\label{eq:62.1}
\end{equation}
for each of the three ladders specified by $\mu=-3,1,2.$ (In what
follows, we deal only with (-) state vectors. The label (-) is to
be understood for these state vectors.) We want to treat each ladder
like a harmonic oscillator ladder, so we define
\begin{align}
\tilde{c}\,|\,\kappa,\mu\,\rangle & =-\frac{\sqrt{2\kappa}}{[8(\mu+3\kappa-1)(\mu+3\kappa-2)(\mu+3\kappa+3)]^{\frac{1}{2}}}c\,|\,\kappa,\mu\,\rangle\nonumber \\
 & =\sqrt{2\kappa}\,|\,\kappa-1,\mu\,\rangle\quad\mathrm{for}\ \mu\neq-3,1,2\label{eq:62.5}
\end{align}
and
\begin{equation}
\tilde{c}\,|\,\kappa,\mu\,\rangle=0\quad\mathrm{for}\ \mu=-3,1,2.\label{eq:62.7}
\end{equation}

As an operator expression, this is
\begin{equation}
\tilde{c}=-\frac{1}{[(H^{(-)}-2)(H^{(-)}-4)(H^{(-)}+6)]^{\frac{1}{2}}}\sqrt{\frac{H^{(-)}-2\mu}{3}}\,c.\label{eq:62.8}
\end{equation}

The algebra is just the modified Heisenberg algebra
\begin{align}
[H^{(-)},\tilde{c}] & =-6\tilde{c},\quad[H^{(-)},\tilde{c}^{\dagger}]=+6\tilde{c}^{\dagger},\nonumber \\
\tilde{c}\,\tilde{c}^{\dagger} & =\frac{1}{3}(H^{(-)}-6-2\mu)+2,\quad\tilde{c}^{\dagger}\tilde{c}=\frac{1}{3}(H^{(-)}-6-2\mu).\label{eq:62.10}
\end{align}

\section{Construction of the Coherent States}

This section is devoted to constructing the appropriate coherent states
for the various ladder operators of Section III. Once these are constructed,
we will examine their physical properties, looking for classical or
non-classical behaviour, in Section V.

Let us point out that ladder operators of the form $a^{3},(a^{\dagger})^{3}$
have been investigated \cite{Castillo-Celeita2016b} for a particular
case of a model related to the fourth Painlevé transcendent \cite{Marquette2016}.
It is believed that our work is the first to treat the coherent states
of the $c,c^{\dagger}$ operators.

There are more general ladder operators of the form $c_{m},c_{m}^{\dagger}$
for $m$ even and greater than or equal to 2 (as well as the trivial
case $m=0$) \cite{Marquette2013b,Marquette2014b}. These lead to
models involving EOPs \cite{Marquette2016} and rational solutions
of the fourth Painlevé transcendent related to the generalized Okamoto
and Hermite polynomials \cite{Marquette2014b}. Our $c,c^{\dagger}$
operators correspond to the case $m=2.$ 

If $\alpha$ is a lowering operator, one of $\{a,c,\tilde{c}\},$
the coherent state $|\,z,\alpha\,\rangle$, for $z$ any complex number,
is defined as the normalized solution of 
\begin{equation}
\alpha\,|\,z,\alpha\,\rangle=z\,|\,z,\alpha\,\rangle.\label{eq:63}
\end{equation}
This is called the Barut-Girardello definition of coherent states
\cite{Barut1971}. There are other definitions involving, for example,
the displacement operator \cite{Perelomov1986}.

The general method of solving such an equation is to first write the
coherent state as a superposition over the appropriate part of the
spectrum of energy (number) eigenvectors. Then Eq.~(\ref{eq:63})
gives a recursion relation which can be solved to give all the coefficients
proportional to the lowest weight coefficient. Finally the state vector
is normalized to unity.

For the $a$ operator acting on the eigenvectors of $H^{(+)},$ the
solution is well known to be 
\begin{equation}
|\,z,a\,\rangle=e^{-|z|^{2}/4}\sum_{\nu=0}^{\infty}|\,\nu\,(+)\,\rangle\frac{(z/\sqrt{2})^{\nu}}{\sqrt{\nu!}},\label{eq:64}
\end{equation}
a superposition over the entire spectrum.

For the $c$ operator acting on the $H^{(-)}$ system, there are three
orthogonal coherent states (three subspaces), for each complex value
of $z,$ with lowest weights $\mu=-3,1,2,$ given by 
\begin{equation}
|\,z,c(\mu)\,\rangle=\frac{1}{\sqrt{F^{(\mu)}(z)}}\sum_{k=0}^{\infty}|\,\mu+3k\,\rangle\frac{z^{k}}{D_{k}^{(\mu)}},\label{eq:68}
\end{equation}
with 
\begin{equation}
F^{(\mu)}(z)=\phantom{|}_{1}F_{3}(1;\frac{\mu+2}{3},\frac{\mu+1}{3},\frac{\mu+6}{3};|z|^{2}/216)\label{eq:69}
\end{equation}
and 
\begin{equation}
D_{k}^{(\mu)}=(-)^{k}6^{3k/2}[(\frac{\mu+2}{3})_{k}(\frac{\mu+1}{3})_{k}(\frac{\mu+6}{3})_{k}]^{\frac{1}{2}}.\label{eq:70}
\end{equation}
(In Eq.~(\ref{eq:68}) and in what follows, the (-) label is to be
understood on the state vectors.) So
\begin{equation}
|\,z,c(\mu)\,\rangle=\frac{1}{\sqrt{\phantom{|}_{1}F_{3}(1;\frac{\mu+2}{3},\frac{\mu+1}{3},\frac{\mu+6}{3};|z|^{2}/216)}}\sum_{k=0}^{\infty}|\,\mu+3k\,\rangle(-1)^{k}\frac{(z/\sqrt{216})^{k}}{[(\frac{\mu+2}{3})_{k}(\frac{\mu+1}{3})_{k}(\frac{\mu+6}{3})_{k}]^{\frac{1}{2}}}.\label{eq:70.05}
\end{equation}

In this expression (see \cite{Gradsteyn1980})
\begin{equation}
(a)_{0}=1,\quad(a)_{k}=a(a+1)\dots(a+k-1)=\frac{\Gamma(a+k)}{\Gamma(a)}\label{eq:65.1}
\end{equation}
and the generalized hypergeometric functions are defined by
\begin{equation}
\phantom{|}_{p}F_{q}(a_{1},\dots,a_{p};b_{1},\dots,b_{q};z)=\sum_{n=0}^{\infty}\frac{z^{n}}{n!}\frac{(a_{1})_{n}\dots(a_{p})_{n}}{(b_{1})_{n}\dots(b_{q})_{n}}.\label{eq:65.2}
\end{equation}

For the $\tilde{c}$ operator acting on the $H^{(-)}$ system, the
three coherent states for each complex value of $z$ are
\begin{equation}
|\,z,\tilde{c}(\mu)\,\rangle=e^{-|z|^{2}/4}\sum_{k=0}^{\infty}|\,\mu+3k\,\rangle\frac{(z/\sqrt{2})^{k}}{\sqrt{k!}}\quad\mathrm{for}\ \mu=-3,1,2.\label{eq:70.0}
\end{equation}

\section{Properties of the Coherent States}

In what follows we consider the properties of the coherent states,
$|\,z,\alpha\,\rangle,$ denoted by $\alpha=a,c(\mu),\tilde{c}(\mu),$
for $\mu=-3,1,2.$ First we find the expectation value of the total
energy in each coherent state as a function of the magnitude, $|z|,$
of the coherent state parameter. Then we calculate and graph the time-dependent
position probability densities for each coherent state over one period.
Finally we construct even and odd Schrödinger cat states for each
of the annihilation operators and plot their position probability
densities over one cycle.

\subsection{Energy expectations}

The expectation of energy $H^{(+)}$ in a coherent state of the operator
$a,$ as a function of the coherent state parameter $z,$ is given
by
\begin{align}
\langle\,z,a\,|\,H^{(+)}\,|\,z,a\,\rangle & =\sum_{\nu=0}^{\infty}\langle\,z,a\,|\,\nu\,(+)\,\rangle(2\ \nu+6)\langle\,\nu\,(+)\,|\,z,a\,\rangle\nonumber \\
 & =6+\langle\,z,a\,|\,a^{\dagger}a\,|\,z,a\,\rangle\nonumber \\
 & =6+|z|^{2},\label{eq:70.3}
\end{align}
using the completeness of the energy eigenvectors, the fact that $a^{\dagger}a$
is diagonal in that basis with eigenvalue $2\nu$ and the defining
relation (\ref{eq:63}) for the coherent states. This is a familiar
result.

We anticipate that the corresponding expressions for the energy expectations
for our other coherent states will differ from this expression. So
we use the energy expectation as one way to characterize the differences
between our ladder operators.

Since the energy eigenvalues in all (-) cases are given by Eq.~(\ref{eq:22}),
we find the expectation of energy as a function of $z$ for $\alpha=c(\mu),\tilde{c}(\mu)$
to be
\begin{equation}
\langle\,z,\alpha\,|\,H^{(-)}\,|\,z,\alpha\,\rangle=6+2\sum_{\nu}|\langle\,\nu\,(-)\,|\,z,\alpha\,\rangle|^{2}\,\nu.\label{eq:71}
\end{equation}
The sum over $\nu$ takes different forms in the different cases.
Note that the results do not depend on the position wavefunctions
of the (-) system.

For $c(\mu)$, $\mu=-3,1,2,$ this is
\begin{align}
\langle\,z,c(\mu)\,|\,H^{(-)}\,|\,z,c(\mu)\,\rangle & =6+2\mu+\frac{1}{(\mu+2)(\mu+1)(\mu+6)}\frac{3|z|^{2}}{4}\frac{\phantom{|}_{1}F_{3}(2;\frac{\mu+2}{3},\frac{\mu+1}{3},\frac{\mu+6}{3};|z|^{2}/216)}{\phantom{|}_{1}F_{3}(1;\frac{\mu+2}{3},\frac{\mu+1}{3},\frac{\mu+6}{3};|z|^{2}/216)},\label{eq:71.5}
\end{align}
using Equations~(\ref{eq:68},\ref{eq:65.1},\ref{eq:65.2}).

Finally, for $\tilde{c}(\mu)$, $\mu=-3,1,2,$ this is
\begin{align}
\langle\,z,\tilde{c}(\mu)\,|\,H^{(-)}\,|\,z,\tilde{c}(\mu)\,\rangle & =6+2\sum_{k=0}^{\infty}e^{-|z|^{2}/2}\frac{(|z|^{2}/2)^{k}}{k!}(\mu+3k)\label{eq:71.6}\\
 & =6+2\mu+3|z|^{2},\label{eq:71.7}
\end{align}
using Equations~(\ref{eq:70.0},\ref{eq:65.1},\ref{eq:65.2}). The
factor of 3 multiplying $|z|^{2},$ compared to Eq.~(\ref{eq:70.3}),
is understandable as the energy levels involved increase in threes.

The results are plotted in Figure 2. We note that the energy expectation
rises far more slowly for the $c(\mu)$ cases than for the cases that
have been constructed to be like the harmonic oscillator.

\begin{figure}
\begin{centering}
\includegraphics[width=16cm]{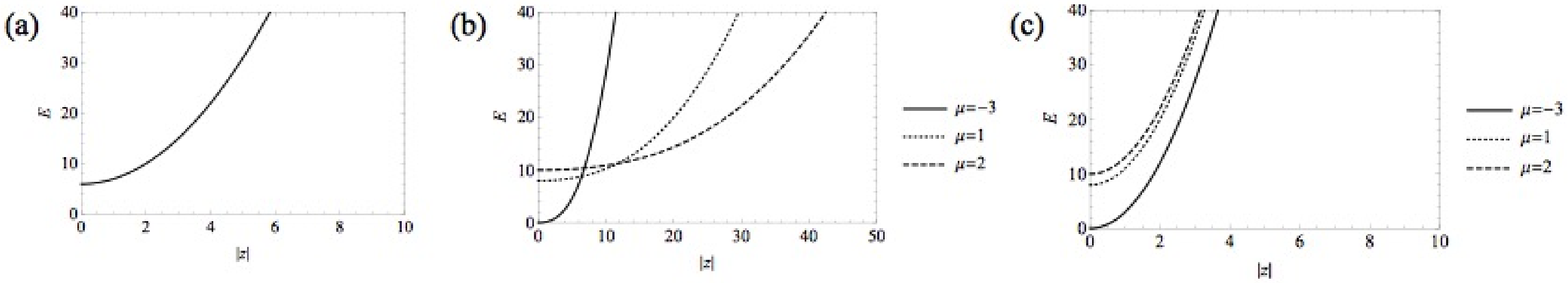}
\par\end{centering}
\caption{Energy expectations as functions of $|z|$ for (a) $a,$ (b) $c(\mu),$
(c) $\tilde{c}(\mu).$}
\end{figure}

\subsection{Time-dependent position probability densities for the coherent states}

The time-dependent position probability density for the (+) system
is
\begin{align}
\rho(x,t;z,a) & =\left|\langle\,x\,|\,e^{-iH^{(+)}t}\,|\,z,a\,\rangle\right|^{2}\nonumber \\
 & =\left|\sum_{\nu=0}^{\infty}\langle\,x\,|\,\nu\,(+)\,\rangle\,e^{-i(6+2\nu)t}\,\langle\,\nu\,(+)\,|\,z,a\,\rangle\right|^{2}\nonumber \\
 & =\left|\sum_{\nu=0}^{\infty}\psi_{\nu}^{(+)}(x)\,e^{-|z|^{2}/4}\frac{(z\,e^{-i2t}/\sqrt{2})^{\nu}}{\sqrt{\nu!}}\right|^{2}.\label{eq:75}
\end{align}
It can be shown \cite{Cohen-Tannoudji1977} that this expression reduces
to
\begin{equation}
\rho(x,t;z,a)=\frac{1}{\sqrt{\pi}}e^{-(x-z\cos2t)^{2}}\label{eq:76}
\end{equation}
for $z$ real.

For the various cases in the (-) system, we have
\begin{align}
\rho(x,t;z,c(\mu)) & =\left|\sum_{k=0}^{\infty}\psi_{\mu+3k}^{(-)}(x)\,(-1)^{k}\frac{1}{\sqrt{\phantom{|}_{1}F_{3}(1;\frac{\mu+2}{3},\frac{\mu+1}{3},\frac{\mu+6}{3};|z|^{2}/216)}}\frac{(z\,e^{-i6t}/\sqrt{216})^{k}}{[(\frac{\mu+2}{3})_{k}(\frac{\mu+1}{3})_{k}(\frac{\mu+6}{3})_{k}]^{\frac{1}{2}}}\right|^{2},\label{eq:78}\\
\rho(x,t;z,\tilde{c}(\mu)) & =\left|\sum_{k=0}^{\infty}\psi_{\mu+3k}^{(-)}(x)\,e^{-|z|^{2}/4}\frac{(z\,e^{-i6t}/\sqrt{2})^{k}}{\sqrt{k!}}\right|^{2}.\label{eq:79}
\end{align}

All of these densities are periodic in time, with period $\pi$ for
$\alpha=a$ and $\pi/3$ for $\alpha=c(\mu),\tilde{c}(\mu).$

We choose to compare all three of the coherent state categories at
the value $z=15.$ We will see in the next subsection that this value
gives interesting behaviour of the Schrödinger cat states.

The results are shown in Figures 3, 4 and 5. In Figure 3 (a) and Figure
5 (a), the $a$ density takes the basic form that is identified as
analogous to classical, with a single wavepacket, like a classical
particle, undergoing oscillatory motion in the potential while retaining
the same width.

\begin{figure}
\begin{centering}
\includegraphics[width=12cm]{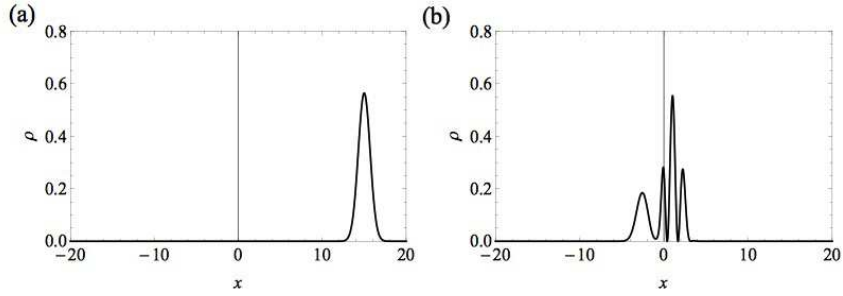}
\par\end{centering}
\caption{Position probability densities at $t=0$ for (a) $a$ and (b) $c(-3)$
with $z=15$ in both cases. }
\end{figure}

The case $c(-3)$ (Figure 3 (b)) was the least classical of the three.
We see several, comparatively indistinct, wavepackets colliding and
oscillating in the potential in our time-dependent simulations.

For the linearised case, $\tilde{c}(-3)$, in Figure 4, we see three
wavepackets, distinct until they collide, when interference fringes
are formed (visible as the blackness in Figure 4 (a).) This is the
closest to classical behaviour, other than the $a$ case, that we
observe. The cases $\mu=1,2$ gave very similar results to Figure
4 so are not shown.

Note that the special case $z=0,$ in all cases, gives the classical
but trivial result of the time-independent probability density of
the lowest weight state.

\begin{figure}
\begin{centering}
\includegraphics[width=16cm]{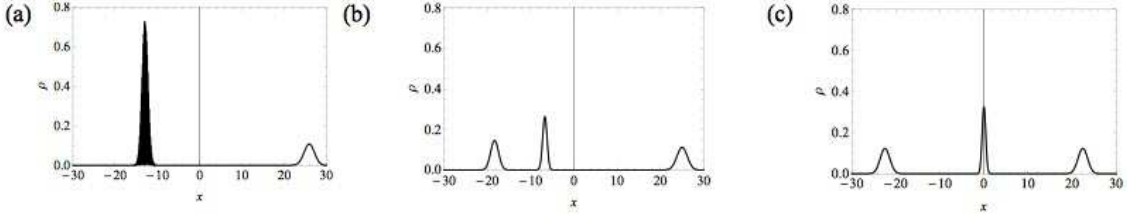}
\par\end{centering}
\caption{Position probability densities for $\tilde{c}(-3)$ for (a) $t=0,$
(b) $t=\pi/24$ and (c) $t=\pi/12$ for $|z|=15$ in all cases.}

\end{figure}

In Figure 5 we show the time dependence of the position probability
densities for the three cases, all for $z=15$, confirming that the
$\tilde{c}(-3)$ case displays more classical behaviour than the $c(-3)$
case. Note in Figure 5 (c) the effect of the narrow, deep part of
the potential on the wavepacket.

\begin{figure}
\begin{centering}
\includegraphics[width=16cm]{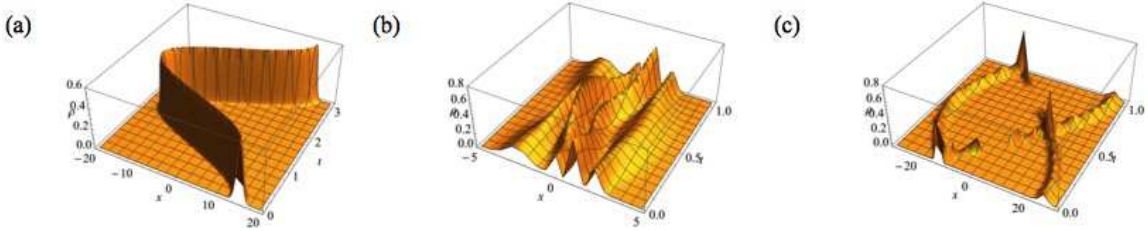}
\par\end{centering}
\caption{Time-dependent position probability densities for (a) $a$ (the harmonic
oscillator), (b) $c(-3)$ and (c) $\tilde{c}(-3),$ all for $z=15.$}

\end{figure}

In Figure 6 we show the probability density at $t=0$ for $c(-3)$
as a function of $z.$ As just noted, the case $z=0$ is trivially
classical, while for increasing $z$ we do not see the well-separated
wavepackets of the case $\tilde{c}(-3).$

\begin{figure}
\begin{centering}
\includegraphics[width=6cm]{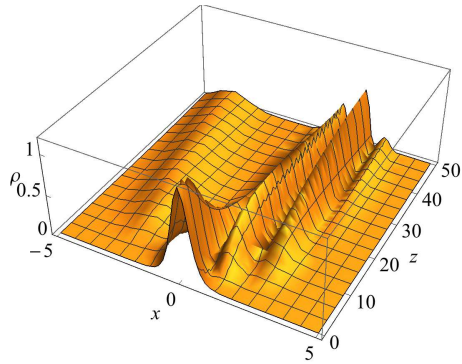}
\par\end{centering}
\caption{Position probability density at $t=0$ for $c(-3)$ as a function
of $z.$}

\end{figure}

\subsection{Time-dependent position probability densities for the even and odd
cat states}

A Schrödinger cat state is a superposition of two macroscopically
distinct states, and is studied in connection with the measurement
problem of quantum mechanics \cite{Mancini1997}. In our case we take
the two states as $|\,+z,\alpha\,\rangle$ and $|\,-z,\alpha\,\rangle,$
for each case of $\alpha.$ We form two normalised superpositions,
one symmetric, the other antisymmetric under the exchange of the two
state vectors:
\begin{align}
|\,z,\alpha,\pm\,\rangle & =\frac{1}{\sqrt{2}}\{|\,+z,\alpha\,\rangle\pm|\,-z,\alpha\,\rangle\}/\sqrt{1+D(|z|,\alpha)}.\label{eq:80}
\end{align}
We calculated the real quantity
\begin{equation}
D(|z|,\alpha)=\langle\,+z,\alpha\,|\,-z,\alpha\,\rangle\label{eq:82.1}
\end{equation}
for all of our cases and found
\begin{align}
D(|z|,\alpha) & =e^{-|z|^{2}}\quad\mathrm{for}\ \alpha=a,\tilde{c}(\mu),\label{eq:83}\\
D(|z|,c(\mu)) & =\frac{\phantom{|}_{1}F_{3}(1;\frac{\mu+2}{3},\frac{\mu+1}{3},\frac{\mu+6}{3};-|z|^{2}/216)}{\phantom{|}_{1}F_{3}(1;\frac{\mu+2}{3},\frac{\mu+1}{3},\frac{\mu+6}{3};|z|^{2}/216)}.\label{eq:85}
\end{align}
These are plotted in Figure 7. The condition for the two states to
be macroscopically distinguishable is
\[
D(|z|,\alpha)\ll1.
\]

\begin{figure}
\begin{centering}
\includegraphics[width=14cm]{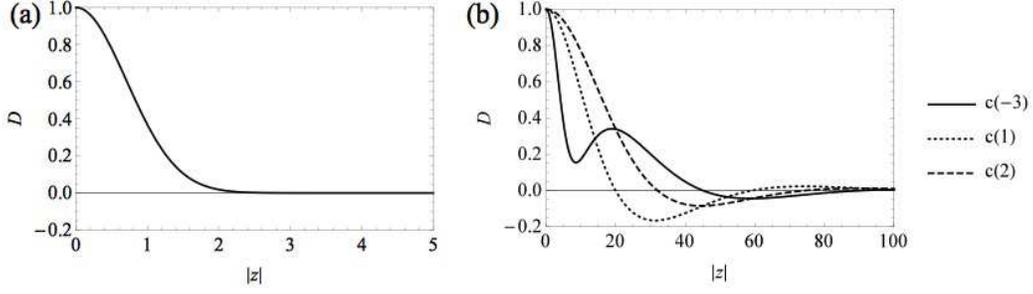}
\par\end{centering}
\caption{The real overlap $D(|z|,\alpha)=\langle\,+z,\alpha\,|\,-z,\alpha\,\rangle$
for (a) $\alpha=a,\tilde{c}(\mu)$ and (b) for $\alpha=c(\mu).$}

\end{figure}

However, we would like to compare all of the cat states at the same
value of $z,$ so we choose $z=15,$ which certainly gives distinguishability
for the $a$ and $\tilde{c}(\mu)$ cases. (We also considered $z=100$
for $c(\mu),$ with that result to follow.)

The cat state results are shown in Figures 8-10.

\begin{figure}
\begin{centering}
\includegraphics[width=12cm]{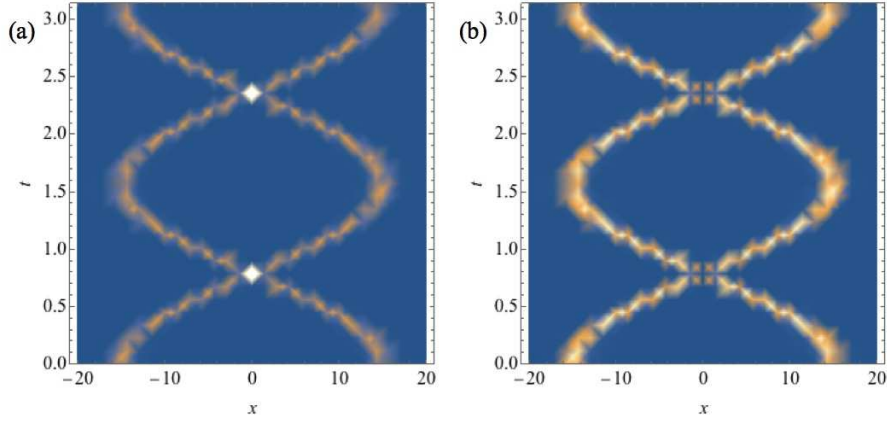}
\par\end{centering}
\caption{Cat state density for $a$ coherent states, (a) even and (b) odd,
for $z=15.$}

\end{figure}

\begin{figure}
\begin{centering}
\includegraphics[width=12cm]{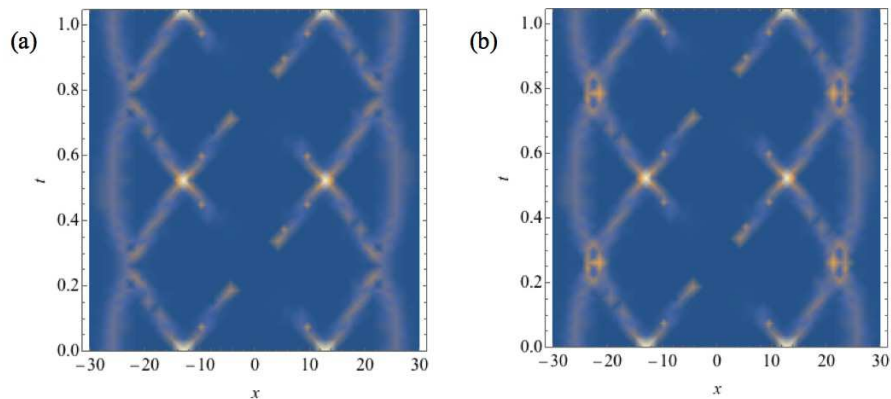}
\par\end{centering}
\caption{Cat state density for $\tilde{c}(-3)$ coherent states, (a) even and
(b) odd, for $z=15.$}
\end{figure}

For the $a$ harmonic oscillator case, the characteristic pattern
shows two wavepackets oscillating $\pi$ out of phase and interfering
at the origin. Note that the period of a cat state will be half the
period of one of its coherent states. The even cat state is a superposition
only over even values of the index $k$, while the odd state contains
only odd $k.$ The position wavefunctions are proportional to the
Hermite polynomials $H_{k}(x),$ which are even for even $k,$ odd
for odd $k.$ Thus we see a nodal line at $x=0$ for the odd cat state.

The $\tilde{c}(-3)$ pattern is, like the densities of Figure 4, the
nearest we see to classical behaviour (other than, of course, the
harmonic oscillator case). We see 6 wavepackets at any time, except
when they collide.

\begin{figure}[h]
\begin{centering}
\includegraphics[width=12cm]{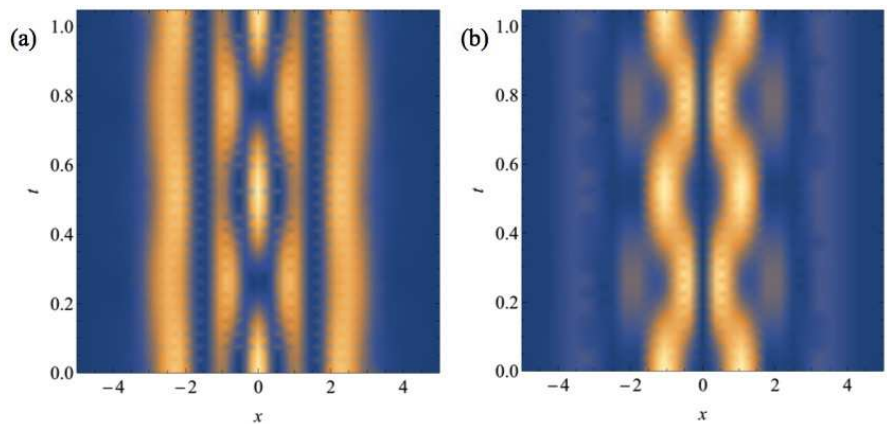}
\par\end{centering}
\caption{Cat state density for $c(-3)$ coherent states, (a) even and (b) odd,
for $z=15.$}
\end{figure}

For the $c(-3)$ case, we see patterns distinctly different from the
earlier cases. We see a nodal line at $x=0$ for the odd case, Figure
10 (b). In all cases, a nodal line at $x=0$ occurs when the appropriate
position wavefunction is an odd function of $x.$ The $\psi_{-3+3k}^{(-)}(x)$
are even for even $k$.

Neither case here can be said to have a classical appearance.

For $c(\mu),$ the cases $\mu=1,2$, even and odd, gave similar patterns
to those seen for $\mu=-3,$ with central nodal lines in the appropriate
places, so are not shown.

As mentioned earlier, we graphed the case $c(-3),$ $z=100.$ We saw
patterns of many peaks and valleys, distinctly non-classical.

\section{Conclusions}

The goal of this paper was to investigate certain properties of the
recently constructed ladder operators, $c$ and $c^{\dagger}$, for
a rationally deformed harmonic oscillator system. This investigation
was to be done by comparing results for coherent states constructed
as eigenstates of the annihilation operator, $c,$ and the more well-known
operator $a$ (for the harmonic oscillator). We also constructed linearized
versions of these operators, $\tilde{c}$ and $\tilde{c}^{\dagger},$
and investigated the same properties for them. We constructed coherent
states $|\,z,c(\mu)\,\rangle$ with lowest weights $\mu=-3,1,2$ and
corresponding coherent states for the linearized operator, $\tilde{c},$
$|\,z,\tilde{c}(\mu)\,\rangle.$ These were compared with the coherent
states $|\,z,a\,\rangle.$

First we plotted the position probability densities at time $t=0$
for the same value of the coherent state parameter, $z,$ for the
various coherent states. The coherent states of the harmonic oscillator
(of the annihilation operator $a$), as is well known, behave in a
very classical way, with a single wavepacket keeping its shape as
it oscillates in the potential. We found that the coherent states
labelled $\tilde{c}(\mu)$, of the linearised operator, display three
well-defined wavepackets oscillating and colliding in the potential.
Certainly at the collision points, when interference fringes are produced,
the behaviour is quantum. But for large sections of time, the wavepackets
are separate, mimicking a classical system. The coherent states labelled
$c(\mu),$ however, show no similar classical regime up to $z=100,$
other than the trivial case $z=0$ of a stationary position probability
distribution.

The even and odd Schrodinger cat states were constructed for each
of the cases and the position probability densities plotted. This
confirmed our conclusion that the operator $c$ leads only to purely
quantum behaviour, with no classical analogue, while the linearised
operator, $\tilde{c}$, has coherent states that can be said to have
a classical analogue.

We have noted that our deformed system is the case $m=2$ of an infinite
class characterized by even $m\geq0$ \cite{Marquette2013b}. In future
work we will investigate coherent states for higher values of $m.$

We also noted that the definition we used for a coherent state as
the complex eigenstate of an annihilation operator is just one of
several different ways to define coherent states in such systems.
These different ways all reduce to the same result for the harmonic
oscillator, but give different results in more general cases. In a
future work we will investigate the Perelemov definition of coherent
states \cite{Perelomov1986}.
\begin{acknowledgments}
IM and YZZ were supported by Australian Research Council Discovery
Projects DP160101376 and DP140101492, respectively. VH acknowledges
the support of research grants from NSERC of Canada. SH receives financial
support from a UQ Research Scholarship.
\end{acknowledgments}

\bibliographystyle{vancouver}

\end{document}